\documentclass[aps,prb,twocolumn,groupedaddress]{revtex4}
\usepackage{graphicx}% Include figure files

\begin{document}
\title{Scaling Behavior of Angular Dependent Resistivity in CeCoIn$_5$: Possible Evidence for d-Wave Density Waves}
\author{T. Hu,$^{1}$ H. Xiao,$^{1}$ T. A. Sayles,$^{2}$ M. B. Maple,$^{2}$ Kazumi Maki,$^{3}$ B. 
$D\acute{o}ra$,$^{4}$ and C. C. Almasan$^{1}$}

%\thanks{}
\affiliation{$^{1}$Department of Physics, Kent State University, Kent, OH 44242, USA}
\affiliation{$^{2}$Department of Physics, University of California at San Diego, La Jalla, 
CA 92903, USA}
\affiliation{$^{3}$Department of Physics and Astronomy, University of Southern California, 
Los Angeles, CA 90089-0484, USA}
\affiliation{$^{4}$Department of Physics, Budapest University of Technology and Economics, 
H-1521 Budapest, Hungary}
\date{\today}
\begin{abstract}

In-plane angular dependent resistivity ADR was measured in the non-Fermi liquid regime of  
CeCoIn$_5$ single crystals at temperatures $T \le 20$ K and in magnetic  fields $H$ up to 14 T. Two  scaling behaviors
were identified in the low-field region where resistivity shows a linear $T$ dependence, separated by a critical angle
$\theta_{c}$  which is determined by the anisotropy of CeCoIn$_5$; i.e.,   ADR depends only on the perpendicular
(parallel) field component  below (above) $\theta_c$. These scaling behaviors and other salient features of ADR are
consistent with d-wave density waves. 
\end{abstract}

\maketitle
The recently discovered heavy fermion compound CeCoIn$_5$ \cite{PetrovicJPCM}  has generated a lot of interest, 
partly due to the many analogies present between this compound and the high transition temperature $T_c$ 
superconductors.  Like the cuprates, CeCoIn$_5$ has a layered crystal structure \cite{Bianchi}, a quasi-two-dimensional
electronic  spectrum \cite{PetrovicJPCM, Hall}, and a superconducting phase appearing at  the border of the
antiferromagnetic phase \cite{Hegger}.  NMR Knight shift  \cite{Kohori, Curro} and heat conductivity \cite {Izawa}
measurements revealed even parity pairing consistent with $d$-wave pairing  symmetry, while specific heat \cite
{PetrovicEL, Kim} and electrical transport \cite {Sidorov} experiments have shown typical non-Fermi liquid  behavior
and have evoked the possibility of a pseudogap. In fact, infrared spectroscopy studies have revealed the development of
a gap in the density of electronic states of CeCoIn$_5$ below $\sim$ 100 K \cite{Singley}. As in the case of cuprates,
all these features suggest the existence  of quantum critical phenomena \cite
{Paglione}. However, the Fermi surface topology is much more  complex than in cuprates, with multiple sheets, which
implies the participation of several bands in the pairing process \cite  {Hall2, Settai, Costa-Quintana}.

In this complex picture, the nature of the normal state in both systems is not  trivial. Until recently, the nature 
of the pseudogap (or the non-Fermi liquid)  phase present in CeCoIn$_5$ has not been addressed.  The giant Nernst
effect found in this material  above $T_c$ \cite{Bel}, however, reopened the problem of the origin  of the pseudogap
state, not only in this compound, but also in the cuprates.  It has recently been shown that the giant Nernst effect in
CeCoIn$_5$ is consistent  with unconventional density waves UDW or d-wave density  waves d-DW \cite{DoraPRB2}. We
recall that in underdoped cuprates, the pseudogap phase has been attributed  to the existence of phase fluctuations of
the superconducting order parameter \cite {Xu, Wang1, Wang2, Capan, Sandu} as well as d-DW \cite{Cappelluti, Benfatto,
Chakravarty}. In particular, the giant Nernst effect observed  in La$_{2-x}$Sr$_x$CuO$_4$, YBa$_2$Cu$_3$O$_{7-\delta}$,
and Bi$_2$Sr$_2$CaCu$_2$O$_{8+\delta}$ has recently been interpreted satisfactorily in terms of d-DW \cite{MakiCAP}. 

Here, we present new information concerning the nature of the pseudogap of CeCoIn$_5$ through angular 
dependent  resistivity ADR. Two different scaling behaviors are revealed in this compound in the non-Fermi liquid
regime for applied magnetic fields less than 5 T. Specifically, for angles below (above) a certain critical angle
$\theta_c$, the field component  perpendicular (parallel) to the $ab$-planes determines the angular-dependent
dissipation.  The critical angle $\theta_c$ is determined by the anisotropy of the material.  These results and the
other salient features of ADR measured in high applied magnetic fields are consistent with d-wave density waves with
Landau quantization of the quasiparticle spectrum.  Our findings bring further understanding of the underlying physics
in the non-Fermi liquid regime of this exotic  compound and could also advance the understanding of the pseudogap state
of the cuprates.   

High quality single crystals of CeCoIn$_5$ were grown using the  flux method. Typical sizes of the crystals 
are 0.5x0.5x0.1mm$^{3}$ with the $c$-axis oriented along the smallest dimension. We determined the out-of-plane
$\rho_c$ and in-plane $\rho_{ab}$ resistivities using the electrical  contact configuration of the flux transformer
geometry, as described in Ref.  \cite {Jiang}. The angular dependent resistivity $\rho_{ab}(\theta)$  was measured by
rotating the single crystal from $H || c-$axis to $H || I || a$-axis, with $\theta$ the angle between $H$ and the
$c$-axis. Here we report only in-plane  resistivity results.

The angular dependent in-plane resistivity of CeCoIn$_5$ single crystals  was measured in the non-Fermi liquid 
regime, for $T\leq 20$ K and in magnetic fields $H$ up to 14 T. Figure 1 is a typical contour plot of resistivity
obtained from $\rho$ vs $\theta$ data measured at different fixed applied magnetic field values and at a temperature of
6 K. A similar topologic contour is obtained over the whole $T$ range investigated. The squares, circles, and triangles
correspond to $n$ = 0.45, 1, and 2/3, respectively,which consistent with previously report \cite{Paglione}, where $n$ is the exponent of the power-law $T$ dependence of
the resistivity; i.e., $\rho(T) =
\rho_0 + AT^{n}$. Note that the contour plot can be divided into three different regions based on the shape of the
topologic plots and the values of $n$. Region I is the rectangular part, for which $n$ = 1. This region
is the so-called low-field region ($H<5$ T) for reasons given later on.
Region II is the elliptical part, for which $n$ = 0.45 and which corresponds to the region around the maximum value of
the $\rho(H)$ curves for different angles. Region III is a large area with $n$ = 2/3, which is outside of regions I and
II at even higher fields. Region I is more interesting since it represents the typical non-Fermi liquid
behavior (for this region $n$ = 1) and displays the two scaling behaviors. Therefore, we will discuss
the anomalous magnetoresistivity behavior in this region in more detail.

The bottom Inset to Fig. 2(a) shows $\rho(\theta)$ measured in a low magnetic field of 2 T. [Data were measured in
region I.]   The resistivity decreases as
$\theta$ increases from
$0^{o}$,  reaches a minimum around
$60^{o}$, and then displays a peak at $90^{o}$. The data of the top Inset to Fig. 2(a) were measured in a high  field of
9 T. [Data were measured in region III.] The resistivity increases with increasing angle and  displays a shoulder,
followed by a peak at
$90^{o}$. 

In the low field region  (region I), a spectacular scaling of these data is achieved by plotting the  resistivity as a
function of the  component of  the applied magnetic field perpendicular to the $ab$-planes, i.e.,
$H_{\perp}=Hcos\theta$, as shown in the main panel  of Fig. 2(a). Note that  all the curves measured in $H \leq 5$ T
overlap for all  $\theta$ values between $0^{o}$ and  a  critical angle $\theta_c \approx 60^o$,  marked by the 
arrows, and deviate from this
$H\cos\theta$ scaling at higher angles.  This value of the critical angle is independent of temperature  and applied
magnetic field for the range investigated.

We employed a second protocol to measure resistivity, in which  we kept the angle between the magnetic field and 
the $c$-axis fixed and scanned the magnetic field. Figure 2(b) shows the  resistivity as a function of the field
component parallel to the $a-$axis ($H_{\parallel}=H\sin \theta$), measured at  6 K for different orientations of $H$.
Note that all the resistivity curves  scale this time with $H\sin\theta$ for $\theta > \theta _c \simeq 60^{o}$.
Therefore,  the resistivity  data follow two scaling laws:
\begin{equation}
\rho(H, \theta)=\left\{ \begin{array}{ll} f_{1}(Hcos\theta) $ for $  \theta \leq \theta_c \approx 60^{o} \\
f_{2}(Hsin\theta)   $ for $  \theta \geq \theta _c \approx 60^{o}. \end{array} \right.
\end{equation}
Figure 1 also clearly shows the presence of the two scaling laws in region I as evidenced by the rectangular
region of the $\rho$ contours.  The presence of these two scaling behaviors indicates that  below (above) the
critical angle, the field  component   perpendicular (parallel) to the
$ab$-planes determines the in-plane dissipation. 
A similar analysis of the angular dependent  resistivity data measured in high magnetic fields (region III)
has shown that the data  are still dominated by the perpendicular (parallel) field  component below (above) the critical
angle, although the two scaling laws are no longer present.  

Next we try to understand the above shape of $\rho(\theta)$ measured in both low and high fields and  the relationship
 between the two scaling behaviors. We plot in the Inset to Fig. 3   the resistivity measured at 6 K in scanning $H$ up
to 14 T at the two fixed angles ($\theta=0^o$ and $\theta=90^o$)  corresponding to the two field orientations
(perpendicular and parallel, respectively, to the $ab$-planes) which seem  to determine the physics below and above
$\theta_c$, respectively.  For the first field orientation ($H\perp ab$ planes), $\rho(H_{\perp})$ increases with
increasing $H_{\perp}$, reaches a maximum around 5 T,  and decreases with further increasing $H_{\perp}$. We define the 
low (high) field
region as the field region for which $H_{\perp} <$ 5 T ($H_{\perp} >$ 5 T),  based on this behavior  of $\rho(H_{\perp})$. 
This
$\rho(H_{\perp})$ dependence gives a qualitative explanation of the $\rho(\theta)$ dependence [shown in the two
Insets to Fig. 2(a)] for $\theta<\theta_c$, i.e., at angles at which the dissipation is dominated by the
perpendicular field component
$H_{\perp}=H\cos\theta$.  Namely, at constant $H$, the resistivity  should decrease (increase) with increasing
$\theta$, i.e., decreasing $H_{\perp}$, for fields lower (higher)  than 5 T as,
indeed, shown by the bottom (top) Inset to Fig. 2(a) for $\theta<\theta_c$.  

For the second field orientation ($H\parallel a$-axis), the $\rho(H_{\parallel})$ data of the Inset to Fig. 3 show 
an  initial increase with increasing $H_{\parallel}$ followed by a tendency to saturation for $H_{\parallel}$
approaching 14 T.  This monotonic $\rho(H_{\parallel})$ dependence explains qualitatively   the  increase in
$\rho(\theta)$ [shown in the two Insets to Fig. 2(a)] for $\theta>\theta_c$, i.e., at angles  at which the
dissipation is dominated by the parallel field component $H_{\parallel}=H\sin\theta$, in both low and high field
regimes. In summary, the $\rho(H_{\perp})$ and $\rho(H_{\parallel})$ dependences of the Inset to Fig. 3 explain the
non-monotonic 
$\rho(\theta)$ dependence  at low fields and the monotonic  $\rho(\theta)$ dependence at high fields, as shown by
bottom and top Inset, respectively, to Fig. 2(a).

Figure 3 shows that the two resistivity  curves of its Inset,  $\rho(H_{\perp})$ and $\rho(H_{\parallel})$, scale   
for $H < 4$ T if  $\rho(H_{\parallel})$ is replaced by $\rho(\gamma^{-1}H_{\parallel})$,  with $\gamma=1.7$. The
physical meaning of $\gamma$ is the anisotropy of the material. The scaling of Fig. 3 implies that the same  physics is
responsible for  the two scaling behaviors displayed by Figs. 1 and 2 for $H < 4$ T, i.e., $f_1(H \cos \theta)$ and
$f_2(H \sin \theta)$, respectively.

The scaling relationship of Fig. 3  also allows one to determine the critical angle $\theta_c$ below (above) which 
the resistivity data  scale when plotted vs the perpendicular (parallel) component of the applied magnetic field 
(Figs. 2(a) and 2(b)).  Recall that Eq. (1) gives: $f_{1}(Hcos\theta_{c}) = f_{2}(Hsin\theta_{c})$. This latter
relationship and the scaling of Fig. 3 [($f_1(H) = f_2(\gamma^{-1}H)$ for $H < $ 4 T)] give $f_1(Hcos\theta_c) =
f_1(\gamma^{-1}Hsin\theta_c)$. Therefore, $\theta_c=tan^{-1}\gamma=59.5^{o}$   with $\gamma = 1.7$, as determined
above. This value of the critical angle is in excellent agreement with the $60^o$ value  obtained experimentally, which
proves the consistency of our analysis of the angular and field dependent data. So, the critical angle is given by
the anisotropy of the material. 

The scaling of the magnetoresistivity curves for $H||c$-axis (for which the Lorentz force is maximum) and for
$H||I||a$-axis (for which the Lorents force is zero) shown in Fig. 3  implies that the spin effect, rather than the
orbital effect, is  responsible for the magnetoresistivity in region I. The fact that in this region the resistivity
is also linear in
$T$, indicates that this linear $T$ dependence is as well a result of spin fluctuations. This conclusion that spin
fluctuations dominate the charge transport in the non-Fermi liquid regime is consistent with recent In-NQR and Co-NMR
experiments\cite {Kawasaki}, which have revealed that the magnetic nature in CeCoIn$_5$ is characterized by strong
antiferromagnetic spin fluctuations in the vecinity of the quantum critical point. 

In the following, we show that all the observed features of ADR of CeCoIn$_5$ in the non-Fermi liquid regime, 
when a magnetic field is rotated  from $H || c-$axis to $H || I || a-$axis with $\theta$ the angle between $H$ and the
$c-$axis, can be consistently discribed in terms of d-wave density waves in a magnetic field.  The unconventional
density wave UDW or d-wave density wave d-DW is a kind of density wave in which the gap function $\Delta$($\vec{k}$) 
vanishes on the line nodes. Therefore, the transition from the  normal state to the UDW is a metal to metal transition
\cite{DoraMPLB}. The UDW  exhibits two characteristics: angular dependent resistivity  and giant Nernst effect. A UDW
in the low  temperature phase of
$\alpha$-(ET)$_{2}$KHg(SCN)$_{4}$ and the  metallic phase of (TMTSF)$_{2}$X with X = PF$_{6}$ and ReO$_4$ has been
identified through ADR \cite{MakiPRL, DoraEL}.

We assume that the electrical conductivity in the non-Fermi liquid regime is given by:
\begin{equation}
\rho(H,\theta)^{-1} = \sum_{n}\sigma_n(Sech^{2}(E_n/2k_BT)),
\end{equation}
where $E_n$ is the energy of all fermionic excitations (holes and particles). As shown below, this new formula 
appears to be more appropriate than the one used earlier \cite{MakiCAP}. In the absence of a magnetic field, we assume
that the quasiparticle  energy spectrum is given by: 
\begin{equation}
E(\vec{k})=\sqrt{\xi^{2}+\Delta^{2}({\vec{k}})},
\end{equation}
where $\xi = \upsilon(k_{\parallel}-k_{F})-(\upsilon' /c)\cos(ck_{z})$ and $\Delta (k) = \Delta \sin (2 \phi)$; 
here $\upsilon$ and $\upsilon'$ are Fermi velocities in the $ab$-plane and along  the $c$-axis, respectively,
$k_{\parallel}$ is the radial wave vector  within the $ab-$plane, $\Delta$ is the maximum value of the d-DW energy gap
$\Delta(k)$, and $\tan\phi = k_{y}/k_{x}$. Here we assume that $\Delta(k)$ has $d_{xy}$-wave symmetry.  In the vicinity
of the nodal points, it is convenient to replace  $\Delta^2 \sin^2 (2 \phi)$ by $\upsilon_2^2k_{\perp}^2$, where
$k_{\perp}$ is perpendicular to $k_{\parallel}$ within the $ab$ plane and $\upsilon_2/\upsilon=\Delta/E_F$.  Then in a
magnetic field which makes an angle $\theta$ with the $c-$axis, the energy spectrum becomes \cite{NersesyanJPCM}: 

\begin{equation}
E^{\pm}_{1n}=\pm\sqrt {2en \upsilon_{2}H( \upsilon |\cos \theta|+\upsilon' \sin \theta)}-\mu
\end{equation}
\begin{equation}
E^{\pm}_{2n}=\pm\sqrt{2en\upsilon_{2}H\ (\upsilon |\cos \theta|-\upsilon' \sin \theta)}-\mu.
\end{equation}
Here $E^{\pm}_{1n}$ and $E^{\pm}_{2n}$ are the two branches of the Landau levels, n = 0, 1, 2,$\dots$, and $\mu$ 
is the chemical potential. We note that in Ref. \cite{MakiCAP} only $E^{\pm}_{1n}$ is considered. The electrical
conductivity is then given by:
\begin{equation}
\rho(H,\theta)^{-1} = \sigma_0+\sum_{i=+,-}\sigma_1(Sech^{2}(x^{i}_{11})+ Sech^{2}(x^{i}_{21})),
\end{equation}
where $x^{\pm}_{11} = E^{\pm}_{11}/2k_{B}T$ and $x^{\pm}_{21} =E^{\pm}_{21}/2k_{B}T$. For simplicity, we take only 
the n = 0  and n = 1 Landau levels.  Equation (6) fits very well the $\rho(\theta)$ data over the whole range of the
applied magnetic field, both in the low and high field regimes. Representative plots of the resistivity data measured
in 4, 8, and 10 T along with the fitting curves are shown in Fig. 4. In these fits we assume that 
\begin{equation}
\sigma_n=A_n\left[1+\left(\frac{H}{H_0}\cos\theta\right)^{2}\right],
\end{equation}
with $n$ = 0, 1.
The fitting parameters determined by fitting the data measured in different magnetic fields are
almost constant, hence field independent, as they should be. Their values are: the anisotropy
$\gamma$ =
$\upsilon/\upsilon'\approx 1.9$,
$\sqrt{\upsilon\upsilon_2} 
\approx 1.3\times 10^{4}$m/s, $\mu= 8.4$ K and  $H_0 = 20$ T. This value of the anisotropy $\gamma$ of 1.9 is very close
to the value of 1.7 required to scale the data shown in Fig. 3. With $\upsilon = 3.3\times 10^{4}$m/s, obtained by
fitting the
$H_{c2}(T)$ curve of  CeCoIn$_{5}$ \cite{Won} and the above value of $\sqrt{\upsilon\upsilon_2}$, we get
$\upsilon_2/\upsilon$ = $\Delta(0)/ E_{F} \approx$ 0.1. Provided that $E_F=450$ K, this gives $\Delta(0)\approx 45$ K,
a value expected from the weak-coupling theory for d-DW with $T_{c}$ = 20 K. \cite{DoraMPLB} This overall consistency
between the obtained and expected values of the fitting paramenters further attests to the appropriateness of the model
used to analyze the experimental data. 

Moreover, the d-DW model given by Eq. (6) also gives the scaling observed experientally at $H <$ 5 T. Figure 5 is 
a plot of the 
$\rho(\theta)$ data (open symbols), the fitting with Eq. (6) (solid line), and the two scalings (dashed lines) 
calculated from Eq. (6) in which
$E^{\pm}_{1n}$ and $E^{\pm}_{2n}$, Eq. (4) and (5), respectively, include only the $H\cos\theta$ or $H\sin\theta$
terms, for $\theta< 60^0$ and $\theta>60^0$, respectively. Note the excellent  agreement between the fitting line and
the two scaling lines. Therefore, even though the model proposed here is quite complex, it gives the two
experimentally observed scaling laws at low magnetic field values.

In summary, comprehensive angular dependent resistivity ADR measurements in the non-Fermi liquid region of 
CeCoIn$_5$ at  $T \leq 20$ K have shown  the presence of two different scaling behaviors in the low-field region
(region I) which involve the $H$ and
$\theta$ dependence and are due to spin fluctuations. The two scaling regions are separated by a critical
angle
$\theta_c$, which is given by the intrinsic  anisotropy. In the scaling region, the resistivity is linear in
$T$. At higher fields, the ADR data are governed by the same physics, even though the two scalings fail. A possible
explanation for this anomalous ADR in terms of
$d$-wave  density waves with Landau quantization of the quasiparticle spectrum was presented. This approach describes
very well the salient features of ADR data: (i) the distinctive angular  dependences observed in the low
($d\rho/dH_{\perp} > 0$) and high ($d\rho/dH_{\perp} < 0$) field regimes follow naturally from Eq. (7); (ii) the low
field scaling behaviors follow from the quasiparticle spectrum given  by Eqs. (4) and (5). In addition, the quasi two
dimensional aspect of d-wave density waves is explored here for the  first time. 
\begin{acknowledgments}
This research was supported by the National Science Foundation under Grant No. DMR-0406471 at KSU and the U. S. 
Department of Energy under Grant No. DE-FG02-04ER46105 at UCSD. The authors acknowledge useful discussions with Dr. V.
Sandu.
\end{acknowledgments}

\section {Figure Caption}

\begin{figure}
\includegraphics[width=0.5\textwidth]{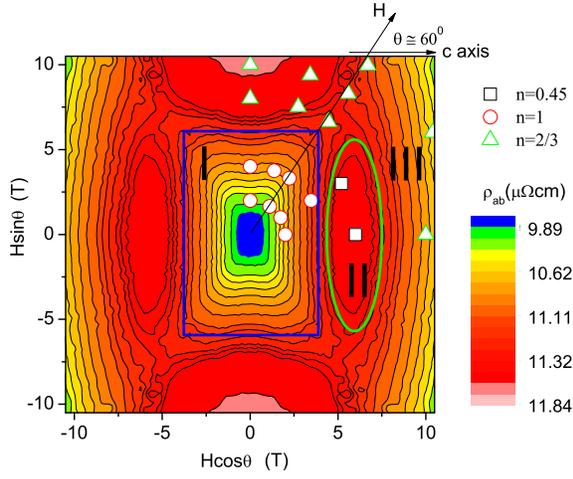}% Here is how to import E_PS art 
\caption{\label{fig:epsart}A contour plot of the in-plane resistivity $\rho$ of CeCoIn$_5$ in the $H\cos\theta-
H\sin\theta$ plane. Three different  regions are found in the contour plot, which are distinguished by different
values of the exponent $n$ of the power-law $T$ dependence of the resistivity; i.e., $\rho = \rho_0 + AT^{n}$. The
$x$-axis of this plot is along the $c-$axis of CeCoIn$_5$ and $\theta_{c}$ is the angle between the diagonal of the
rectangle in region I and the $x$-axis ($c-$axis). }
\end{figure}

\begin{figure}
\includegraphics[width=0.5\textwidth]{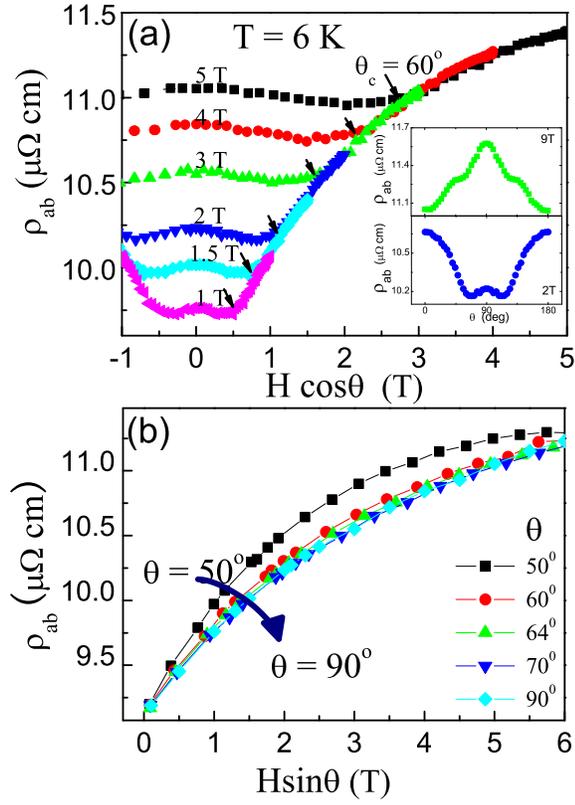}% Here is how to import E_PS art 
\caption{\label{fig:epsart} Figure 2(a) Scaling of resistivity $\rho$ vs the component of the applied magnetic field $H$ 
perpendicular to the $ab$-planes, Hcos$\theta$, measured at a temperature $T$ of 6 K and for 1 T $\leq H\leq$ 5 T. The
arrows mark the points at which the curves start to deviate from each other. Insets: $\rho$ vs the angle $\theta$
measured at in an applied mahnetic field of 2 T and 9 T. 2(b) Resistivity $\rho$ vs the component of the applied
magnetic field $H$ parallel to the $ab$-planes, Hsin$\theta$, measured at a temperature $T$ of 6 K in scanning $H$ at fixed orientation $50^0 \leq \theta \leq 90^0$. The $\rho$ vs $H\sin\theta$ data scale for $60^0 \leq \theta\leq 90^0$. }
\end{figure}

\begin{figure}
\includegraphics[width=0.5\textwidth]{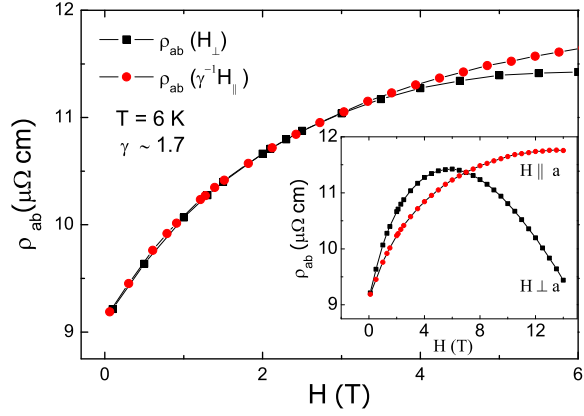}% Here is how to import E_PS art 
\caption{\label{fig:epsart}Figure 3. In region I,  scaling of resistivities $\rho(H_{\perp})$ and $\rho(\gamma^{-1}
H_{\parallel})$ when plotted vs the component of the magnetic field perpendicular to the $ab$-planes, $H_{\perp}$, and
parallel to the $ab$-planes, $H_{\parallel}$, times the inverse anisotropy $\gamma^{-1}$, respectively. Inset: Plots of
$\rho(H_{\perp})$ and $\rho(H_{\parallel})$ curves measured in magnetic fields up to 14 T. }
\end{figure}

\begin{figure}
\includegraphics[width=0.5\textwidth]{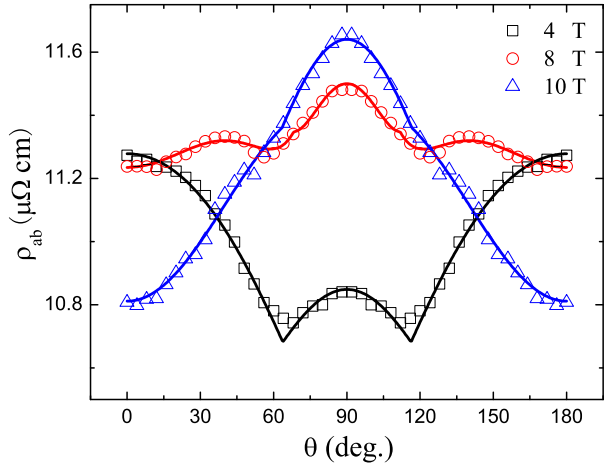}% Here is how to import E_PS art 
\caption{\label{fig:epsart}Figure 4. Angular ($\theta$) dependent resistivity $\rho$ measured at 6 K in a magnetic field of 
4 T, 8 T and  10 T. The solid lines are fits of Eq. (6) to the data.  }
\end{figure}

\begin{figure}
\includegraphics[width=0.5\textwidth]{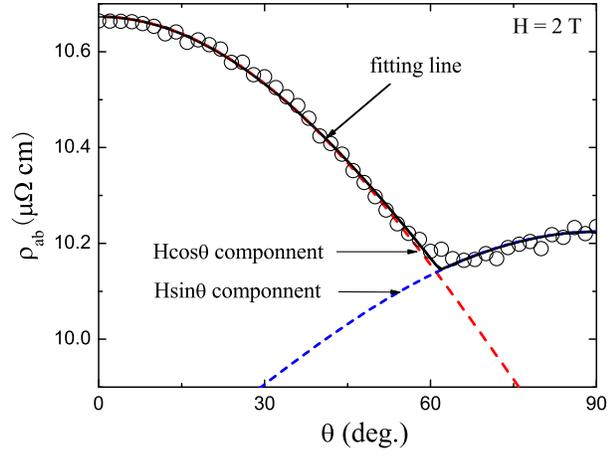}% Here is how to import E_PS art 
\caption{\label{fig:epsart}Figure 5.  Angular ($\theta$) dependence of the resistivity $\rho$ measured at 6 K in a magnetic 
field H = 2 T (open symbols).  The solid line is the fit of Eq. (6) to the data. The dashed lines are the $H\cos\theta$
and $H\sin\theta$ scalings laws.  }
\end{figure}

\end{document}